\documentclass[10pt,aps,twocolumn,prc,superscriptaddress,noshowpacs,nofootinbib,noshowkeys,floatfix]{revtex4}
\usepackage[dvips]{graphics,graphicx}
\usepackage[colorlinks=true,linktocpage=true,linkcolor=blue,citecolor=blue,urlcolor=blue]{hyperref}
\usepackage[usenames,dvipsnames]{color}
\usepackage{amsmath, amssymb}
\usepackage{multirow}
\usepackage{longtable}
\usepackage{bm}
\usepackage{color}
\usepackage{pbox}
\usepackage[normalem]{ulem}  

\renewcommand\sout{\bgroup \color{blue} \ULdepth=-.5ex \ULset}
\usepackage{lipsum}
\usepackage{lipsum}
\usepackage{color,soul}
\setul{0.5ex}{0.3ex}
\definecolor{Red}{rgb}{1,0.0,0.0}
\setulcolor{Red}
\begin{document}

\preprint{}

\title{Analysis of quarkonium polarization in proton-proton (p-p) collisions at LHC using PYTHIA model}

\author{Deekshit Kumar}
\email{deekshitkumarvecc@gmail.com}
\affiliation{Variable Energy Cyclotron Centre, 1/AF Bidhan Nagar, Kolkata 700 064, India}
\affiliation{DPBS College Anupshahr, Bulandshahr,  203390, U.P, India}
         
\author{Ekata Nandy}
\email{ekata@vecc.gov.in}
\affiliation{Variable Energy Cyclotron Centre , 1/AF Bidhannagar, Kolkata, 700064, West Bengal, India} 

\author{Biswarup Paul}
\email{biswarup.babu@gmail.com}
\affiliation{Bali Ram Bhagat College, Samastipur, 848101, Bihar, India}

\author{Subikash Choudhury }
\email{subikash.choudhury@gmail.com}
\affiliation{Nuclear and Particle Physics Research Centre, Department of Physics, Jadavpur University, Kolkata, 700032, West Bengal, India }

\author{Tinku Sinha }
\email{tinkusinhasarkar2022@gmail.com}
\affiliation{Saha Institute of Nuclear Physics , Kolkata, 700064, West Bengal, India }

\author{Partha Pratim Bhaduri}
\email{partha.bhaduri@vecc.gov.in}
\affiliation{Variable Energy Cyclotron Centre, 1/AF Bidhan Nagar, Kolkata 700 064, India}

\date{\today}

\begin{abstract}
 The measurement of polarization serves as an important probe to investigate the production mechanism of quarkonia, the bound state of heavy quark anti-quark (charm or bottom) pairs, in hadronic collisions. In experimental invesigations, the polarization is usually measured by analyzing the anisotropies in the angular distribution of the muons originating from the decay of the quarkonium state. In the present article, we study the charmonia ($J/\psi$) and bottomonia ($\Upsilon(1S)$) polarization at $\sqrt{s} =7 $ and 13 TeV in proton-proton(p-p) collisions at LHC using Monte Carlo (MC) event generator model PYTHIA8, which is based on perturbative QCD. The transverse momentum ($p_{T}$) differential distribution  has been calculated at forward rapidity ($2.5 < y_{\mu\mu} < 4.0$) and the polarization parameters are estimated in Helicity and Collins-Sooper reference frames. In addition, to mimic realistic experimental conditions, we have incorporated, in PYTHIA simulations, effects like detector inefficiencies and muon momentum smearing. These contributions alter the polarization parameters, introducing an artificial degree of polarization, if not properly corrected for. The simulation results have been compared with the recent ALICE measurements for quarkonia polarization in p-p collisions at LHC energy regime.

\end{abstract}

\pacs{25.75.-q,12.38.Mh}
\maketitle


\section{Introduction}

The experimental investigations of the decay distributions of the vector particles offer a unique tool to examine the fundamental theories. In particular, quarkonia polarization measurements are expected to provide key information for the understanding of quantum chromodynamics (QCD), the fundamental theory of strong interaction. The non-relativistic bound states of heavy quark anti-quark pair, stable under strong interaction, are collectively called quarkonia. Even though discovered more than five decades ago, their production in energetic hadronic collisions is not completely understood from the first principle QCD calculations~\cite{DK2001}. Over the years several QCD inspired theoretical models have been developed in the literature to explain the multiscale dynamics of quarkonium production in nuclear collisions. Majority of the theoretical models usually describe the quarkonium production in the hadronic collisions as a factorizable two step process. The first stage is the production of a heavy quark $\&$ anti-quark ($Q\bar{Q}$) pair via partonic hard scattering and computable within perturbative QCD (pQCD). This is followed by the evolution of the $Q\bar{Q}$ pair transforming into a color neutral physical resonance state. The process is non-perturbative in nature and thus model dependent. The underlying assumption of the color-singlet model (CSM)~\cite{ DK2002,DK2003,DK2004} is that the nascent $Q\bar{Q}$ pair bears the same spin ($S$), angular momentum ($L$) and colour quantum numbers of the physical quarkonium state since its production in hard partonic collisions. No transition involving $L/S$-change takes place during the formation of the bound-state. $J/\psi$ and $\psi(2S)$ production cross sections as measured by Fermilab E789 collaboration~\cite{ DK2097}  were found to exceed the model predictions by factor of 7 and 25 respectively. Subsequently the CDF collaboration~\cite{ DK2098} at Tevatron measured around 50 times larger $J/\psi$ and $\psi(2S)$ production cross sections in $\sqrt{s} = 1.8$ TeV p-p collisions, than that calculated by CSM. To overcome the significantly large discrepancy between measurements and model calculations, the non-relativistic quantum chromodynamics (NRQCD) factorization approach~\cite{DK2005,DK2006} has been proposed in literature. Unlike CSM where only color singlet quark pairs can form the bound state mesons, the theoretical construct of NRQCD includes resonance formation  from quark pairs originally produced in colour-octet states as well. In this effective field theory calculation, the long-distance matrix elements (LDMEs), factorized from the parton-level contributions, are introduced as model parameters that convert the pre-resonant coloured $Q\bar{Q}$ pairs into physical bound states, possibly changing 
$L$ and/or $S$ as required. By construction LDMEs are universal (i.e. process-independent) and constant (i.e. independent of the $Q\bar{Q}$ momentum) within the NRQCD framework. Their values are fixed by comparison to experimental data. Production cross section of different quarkonia states as computed within NRQCD factorization approach has been found to be in reasonable agreement with the experimental measurements at Tevatron ~\cite{DK2018,DK2019,DK2020}, RHIC ~\cite{DK2015,DK2016,DK2017} and LHC~\cite{DK2007,DK2008,DK2009,DK2010,DK2011,DK2012,DK2013,DK2014}. Many other theoretical models like color evaporation model (CEM)~\cite{DK2027,DK2028,DK2029}, and $k_{T}$ factorization approach~\cite{DK2030,DK2031} among the others are also in vogue to explain the observed quarkonium production in particle collisions. 

However, for a complete understanding of the quarkonium production in hadronic collisions, the polarization of these vector meson states also demands a careful investigation, both theoretically as well as experimentally. Polarization referes to the degree of spin alignment of these vector bosons relative to a chosen axis, generally referred as quantization axis. Different theoretical models predict different polarization patterns, in a given frame of reference. The color-octet model of NRQCD predicts large transverse polarization of $J/\psi$ mesons at high momenta due to their predominant production from gluon fragmentation preserving natural spin orientation~\cite{DK2091, DK2092}. The CSM model on the other hand predicts a strong transverse ploarization at the leading order, which changes to a strong longitudinal polarization once the next-to-leading order (NLO)  ~\cite{DK2088, DK2089} contributions are included in the model calulations. Depending on the polarization frame, either zero or slightly transverse polarization at high $p_{T}$ and a small longitudinal polarization at low $p_{T}$ is predicted by the improved color evaporation model ~\cite{DK2090} .

In experimental investigations, quarkonium polarization is primarily investigated via the dilepton decay channel. The angular distributions of decay products are analyzed to determine the polarization of the quarkonium states. These distributions can be analyzed in different reference frames such as Helicity frame (HE) and Collins–Soper (CS) frame each defined by specific orientation of the quantization axis. Measurements usually involve the determination of the polarization parameters ($\lambda_{\theta}, \lambda_{\phi}$ and $\lambda_{\theta\phi}$) which quantify the degree of polarization in different directions. In the experimental sector, quarkonium polarization measurements have been performed by E866~\cite{DK2086, DK2087} and HERA-B~\cite{DK2053}  collaborations at fixed target facilities and in collider experiments at Tevatron~\cite{DK2038,DK2039,DK2040,DK2041,DK2042}, RHIC~\cite{DK2035,DK2036,DK2037} and LHC~\cite{DK2043,DK2044,DK2045,DK2046,DK2047,DK2048,DK2049,DK2050,DK2051,DK2052}. The ALICE collaboration at LHC has reported polarization parameters, within uncertainty consistent with zero, for inclusive $J/\psi$ mesons in $\sqrt{s}=7, 8$ TeV p-p collisions ~\cite{DK2093, DK2094} and preliminary result for $\Upsilon(1S)$ mesons in $\sqrt{s} = 13$ TeV p-p collisions, in the forward rapidity ($2.5 < y < 4.0$) region. The LHCb collaboration, on the other hand reports a small longitudinal polarization for prompt $J/\psi$ mesons in the HE frame for pp collisions at $\sqrt{s}$ = 7 TeV~\cite{DK2095}. A similar trend has been observed in the ALICE measurement for the inclusive $J/\psi$ sample but with large systematic uncertainties. Extraction of polarization parameters for $J/\psi$ and $\psi(2S)$ states using ALICE Run2 data in $\sqrt{s}=13$ TeV p-p is underway. The persistent inconsistency between the kinematic dependence of the measured polarization parameters and their model dependent estimates is popularly known as $J/\psi$ polarization puzzle~\cite{DK2096}.

Recently in Ref.~\cite{DK2099} an attempt has been made to phenomenologically study the polarization of $J/\psi$ and $\psi(2S)$ mesons produced in $\sqrt{s} = 7, 8 $ and 13 TeV p-p collisions at LHC. The \texttt{PYTHIA8} event generator has been employed to simulate the charmonia production and the corresponding $\lambda$-polarization parameters ($\lambda_{\theta}, \lambda_{\phi}$ and $\lambda_{\theta\phi}$) have been studied as functions of $p_{T}$, $y$ and charged particle multiplicity ($N_{ch}$) and non-zero values for polarization parameters extracted from \texttt{PYTHIA}-generated events are reported. However, by construction, quarkonia produced in \texttt{PYTHIA}~\cite{DK2084,DK2084a} are unpolarized and decay isotropically and hence should result in null polarization parameters. To investigate this apparent discrepancy, we have reanalyzed the $\lambda$-polarization parameters using \texttt{PYTHIA} simulations, incorporating detector inefficiencies and muon momentum smearing to emulate realistic experimental conditions. This allows us to assess whether such detector effects can induce artificial polarization signals and to evaluate the extent to which these biases can be corrected after accounting for experimental effects.

The rest of the article is organized as follows; a brief description of the \texttt{PYTHIA8} event generator is given in Section II. Analysis methodology for extraction of the polarization parameters from the angular distribution of the decay muons is depicted in section III, while section IV presents the results of the present simulation investigations. Finally, in section V we summarize our results and conclusions.

\begin{figure}
    \includegraphics[width=1.00\linewidth]{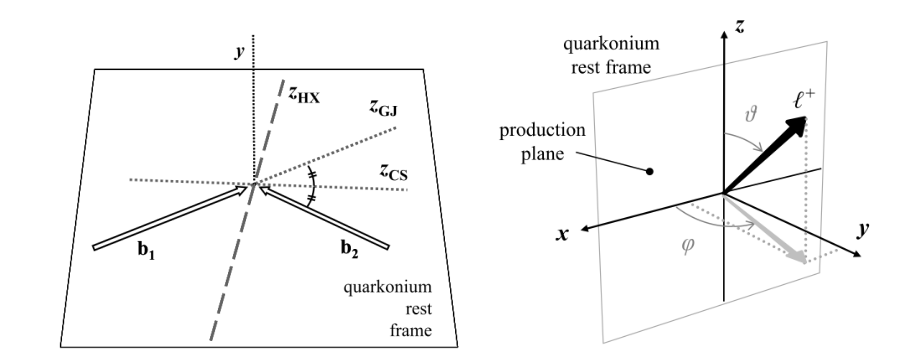}
	\caption{Left panel: Definition of different reference axes for polarization study in the quarkonium rest frame, with HE =  Helicity frame, CS = Collins-Soper frame, and GJ = Gottfried-Jackson frame. Right panel: definition of the decay lepton angles in the quarkonium rest frame. For the present analysis we have taken only HE and CS frames. The figure has been adopted from~\cite{DK2080}.}
	\label{cs_he}		
\end{figure}


\section{\bf{Event Production in PYTHIA8}}

\texttt{PYTHIA8} is a QCD-inspired Monte Carlo event generator widely used to simulate complex multiparticle final states in high-energy hadronic collisions ~\cite{DK_pythia1}. It begins with the elementary hard scattering of two partons, one from each incoming hadron, whose cross sections are computed using perturbative QCD (pQCD) at leading order (LO). These hard processes typically involve \(2 \to 2\) scatterings of light quarks and gluons, as well as heavy-flavor production such as charm and bottom quarks. To prevent the divergence of cross sections as the transverse momentum \(p_{T} \to 0\), suitable phase space cuts or regularization schemes are applied. 

Following the hard scattering, \texttt{PYTHIA8} models initial- and final-state radiation through parton showers, which simulate the QCD evolution of partons by iteratively generating emissions of softer and collinear gluons and quark-antiquark pairs. 

Subsequently, hadronization in \texttt{PYTHIA8} is modeled using the Lund string fragmentation model, where the color field between the partons are represented by longitudinally stretched strings that breaks into hadrons through the creation of quark-antiquark pairs ~\cite{DK_pythia2}.

In addition, \texttt{PYTHIA8} incorporates multiple parton interactions (MPI), accounting for the possibility of several semi-hard scatterings occurring within a single had -ron-hadron collision ~\cite{DK_pythia3, DK_pythia4}. MPI contributes significantly to the underlying event activity and affects the final-state particle multiplicities and kinematic distributions.

An important feature of \texttt{PYTHIA8} is its modeling of \textit{color reconnection} (CR), a non-perturbative process in which color strings originating from different parton interactions are reconnected before hadronization to minimize the total string length ~\cite{DK_pythia5}. Since particle production occurs through the breaking of these strings, reducing the overall string length leads to fewer particles being produced but with higher average momentum. This mechanism plays a key role in explaining the observed hardening of transverse momentum (\(p_{T}\)) spectra as the event multiplicity increases.

Quarkonia production in \texttt{PYTHIA8} is implemented primarily through gluon fusion (\(gg \to Q\bar{Q}\)) and/or quark-antiquark annihilation (\(q\bar{q} \to Q\bar{Q}\)) processes, employing perturbative QCD (pQCD) within the framework of NRQCD~\cite{DK_pythia6}. Quarkonia are produced in both color-singlet (CS) and color-octet (CO) states. However, for the transition from the CO state to the physical quarkonium,  an additional soft gluon is emitted. The overall quarkonium production cross section is calculated in the NRQCD framework by factorizing it into perturbatively computed short-distance coefficients for heavy quark-antiquark pair production in specific color and angular momentum states from initial hard scattering ~\cite{DK_pythia7, DK_pythia8, DK_pythia9}, multiplied by non-perturbative LDMEs that describe the hadronization into physical quarkonia~\cite{DK_pythia10}.

The results presented in this paper are mainly based on the \texttt{PYTHIA8} tune 4C with MPI and CR enabled. To avoid divergence in QCD processes we consider a minimum p$_{T}$ cut of 0.5 GeV. For Quarkonium  production we use the flag \texttt{Charmonium:all=on} and \texttt{Bottomonium: all=on} which allows production of prompt and non-prompt J/$\psi$ and $\Upsilon (1S)$ in CS and CO states. The analysis has been carried out for pp collisions at $\sqrt{s}=$ 7 and 13 TeV.
\texttt{PYTHIA8} model is believed to provide a comprehensive and realistic simulation of hadronic collisions, suitable for both theoretical studies and experimental analyses.

\begin{figure}[ht]
	\centering
	\includegraphics[width=1.00\linewidth]{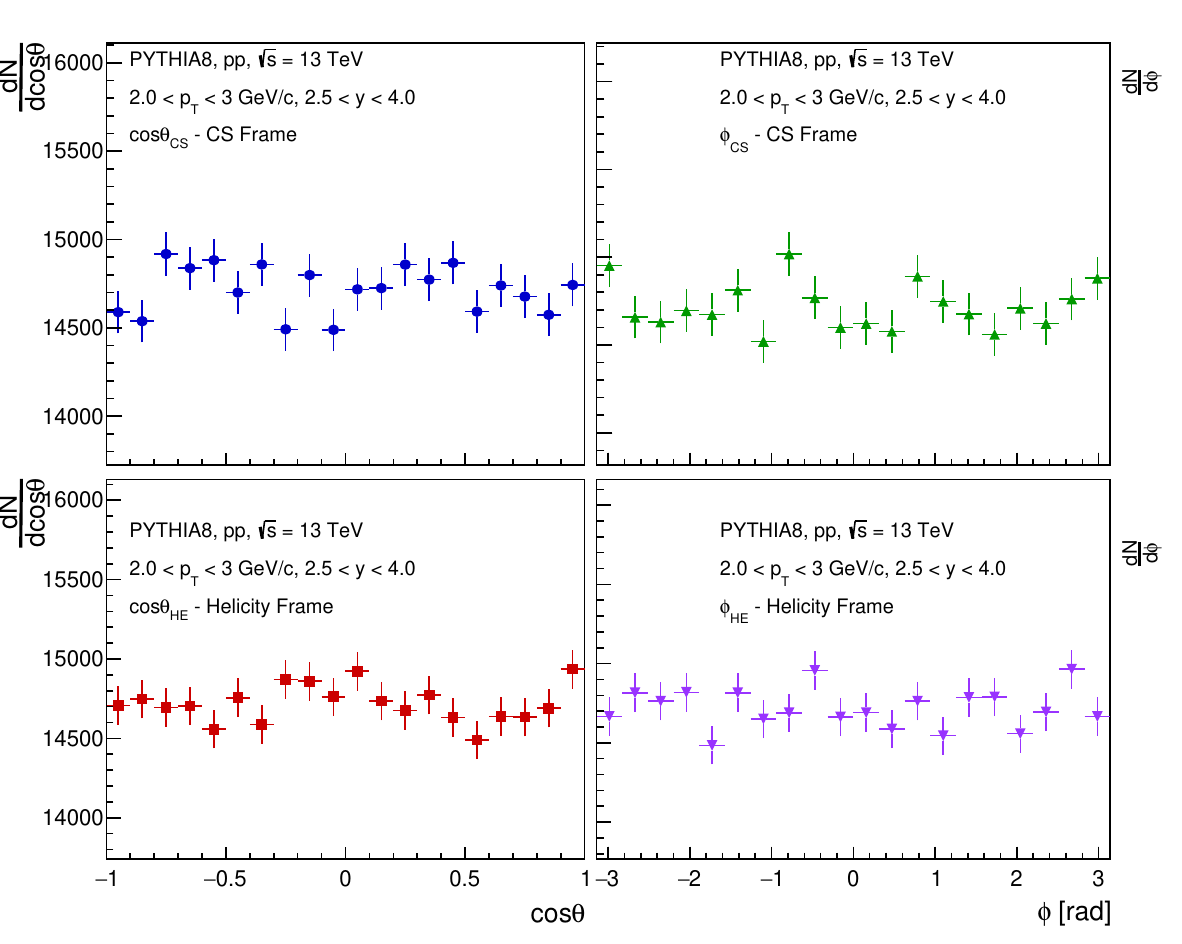}
	\caption{The distributions of the cosine of the polar angle ($\theta$) (left panel) and the azimuthal angle ($\phi$) (right panel) of the di-muons decayed from $J/\psi$ in p-p collisions at $\sqrt{s}$= 13 TeV from \texttt{PYTHIA8} in both the HE  and CS  reference frames.}
	\label{Angular_plotvsPt}
\end{figure}

\section{Analysis Methodology}

\begin{figure}
    \includegraphics[width=1.00\linewidth]{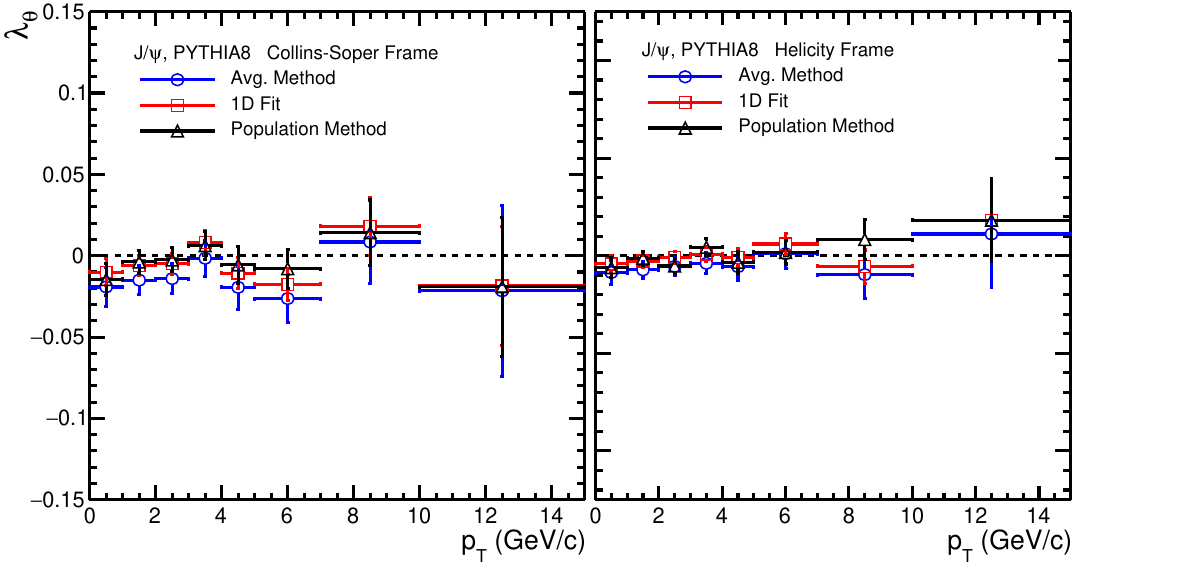}
	\includegraphics[width=1.00\linewidth]{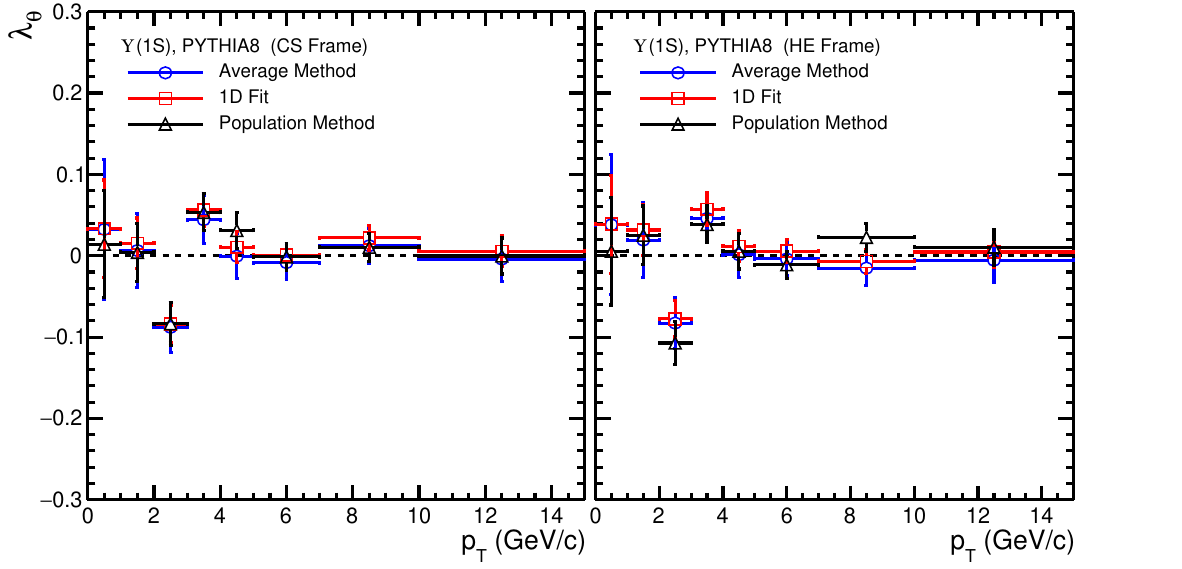}
	\caption{ The polarization parameter ($\lambda_{\theta}$) as a function of transverse momentum ($p_{T}$) for $J/\psi$ and $\Upsilon(1S)$ in p–p collisions at $\sqrt{s} = 13$ TeV in the forward rapidity region. Results are shown in both the HE  and CS reference frames. Three extraction methods are compared: the average method, the 1D angular fit method, and the population method.}
	\label{lam_pt_jpsi_upsilon}		
\end{figure}	

The polarization describes the orientation of the particle spin with respect to a chosen axis known as quantization axis in a given reference frame. For the unstable quarkonium states ($J^{PC} = 1^{--}$) under consideration, the geometrical shape of the angular distribution of the daughters in the two body decay ($^{3}S_{1} \rightarrow \mu^{+}\mu^{-}$), emitted back to back in quarkonium rest frame, reflects polarization of the parent quarkonium state. Quarkonia if produced unpolarized results in a spherically symmetric distribution whereas anisotropies in the angular distribution would indicate polarized production. Measurement of the angular distribution requires a specific co-ordinate system to be defined relative to which the three momentum of one of the two decay products (usually $\mu^{+}$) is expressed in spherical coordinates. The axes of the coordinate system are chosen with respect to the physical reference provided by the directions of the two colliding beams as seen from the quarkonium rest frame, for inclusive quarkonium measurements. The choice of the polar axis is however not unique, and three different conventions are usually followed for the orientation of the polar axis, leading to three different frames of reference, in analysis of the quarkonium decay distributions. The HE aligns the z-axis with the momentum direction of $J/\psi$ (in the centre-of-mass of the two colliding beams), conversely, the CS aligns the z-axis along the bisector of the angle between the two incident proton beams and in Gottfried-Jackson frame (GJ) aligns the z-axis with the direction of the momentum of one of the colliding beams, in the quarkonia rest frame. Figure \ref{cs_he} adopted from Ref.~\cite{DK2080} schematically shows the orientation of the polar axis in three reference frames (left panel) and definition of the decay angles in the quarkonium rest frame. The quantization axis of the GJ frame lies between the CS reference frame and the HE reference frames and thus represents an intermediate situation~\cite{DK2069,DK2080}. Hence in line with the previous analyses, we present our results only in HE and CS frames. In either of the reference frames the two dimensional decay angular distribution $W(\theta, \phi)$ can be parameterized as~\cite{DK2069}:
\begin{eqnarray}
	W(cos\theta, \phi) \propto \frac{1}{3 + \lambda_{\theta}} (1  +\lambda_{\theta} cos^{2}\theta \nonumber\\ + \lambda_{\phi} sin^{2}\theta cos2\phi \nonumber\\ + \lambda_{\theta \phi} sin2\theta cos\phi )
	\label{Angular_distn}
\end{eqnarray}
where $\theta$ and $\phi$ respectively denote the polar and azimuthal angle of the $\mu^{+}$ measured in the quarkonim rest frame and 
$\lambda_{\theta}$, $\lambda_{\phi}$ and $\lambda_{\theta \phi}$ are the $J/\psi$  polarization parameters. The various combinations of these parameters corresponding to the elements of the spin density matrix of $J/\psi$ production can give us information about the polarization of the particle; the values of  ($\lambda_{\theta}$, $\lambda_{\phi}$ , $\lambda_{\theta \phi}$ )  = (0,0,0) defines no polarization, (-1,0,0) defines complete longitudinal polarization and (+1,0,0) defines complete transverse polarization.

Instead of fitting the above 2-D distribution and simultaneously extracting all three $\lambda$ parameters, experimental analyses usually adopt a one-dimensional (1-D) approach and fit the angular distributions obtained by integrating Eq.~\ref{Angular_distn} over cos$\theta$ and $\phi$. The resultant 1-D distribution functions $W(\cos\theta),  W(\phi) \&  W(\tilde{\phi})$ appear as:
\begin{eqnarray}
	W(cos\theta) \propto \frac{1}{3 + \lambda_{\theta}} (1  + \lambda_{\theta} cos^{2}\theta)
	\label{Angular_distn2}
\end{eqnarray}
\begin{eqnarray}
	W(\phi) \propto (1 +   \frac{2 \lambda_{\phi} }{3 +\lambda_{\theta}}   cos2\phi)
	\label{Angular_distn3}
\end{eqnarray}
\begin{eqnarray}
	W(\tilde{\phi} )\propto (1 +  \frac{\sqrt{2} \lambda_{\theta \phi} }{3 +\lambda_{\theta}}  cos\tilde{\phi})
	\label{Angular_distn4}
\end{eqnarray}
The variable $\tilde{\phi}$ is built to estimate  $\lambda_{\theta \phi} $ and can be defined as;
\begin{eqnarray}
	\tilde{\phi} = \phi - \frac{3}{4} \pi, (cos\theta < 0)
	\label{phitilde}
\end{eqnarray}
\begin{eqnarray}
	\tilde{\phi}= \phi - \frac{1}{4}\pi,  ( cos\theta > 0 )
	\label{phitilde2}
\end{eqnarray}

This method allows simultaneous (for a 2-D fit) or sequential (for 1-D fit) extraction of all parameters ($\lambda_{\theta}$, $\lambda_{\phi}$, $\lambda_{\theta\phi}$) but it is very sensitive to the choice of fit range, binning, and statistical fluctuations particularly in low-statistics bins. Instead of fitting the angular distributions (1-D fit in the present case), alternative prescriptions such as `population counting method' and `average angular distribution method' have also been proposed for extraction  of polarization parameters. The two methods are described below.

\begin{enumerate}
	
	\item Population counting method \\
	
	In the \textit{population counting} or \textit{bin-by-bin ratio} method, the $(\cos\theta, \phi)$ space is divided into discrete bins, and the number of signal events in each bin is counted. 
	The ratios of yields between different angular bins are then used to infer the polarization parameters. 
	This method provides an intuitive way to visualize the angular dependence of the decay distribution and serves as an important cross-check of the fitting procedure.

	The following Eqs. [\ref{population1}, \ref{population2} $\&$ \ref{population3}] are employed for this method, where $N$ denotes population of decay muons:
	\begin{equation}
		\frac{N(|\cos \theta| > \frac{1}{2}) - N(|\cos \theta| < \frac{1}{2})}{N(|\cos \theta| > \frac{1}{2}) + N(|\cos \theta| < \frac{1}{2})} = \frac{3}{4}  \frac{3 \lambda_{\theta}}{3+ \lambda_{\theta}}
		\label{population1}
	\end{equation}
	\begin{equation}
		\frac{N((\cos2 \phi) > 0 ) - N((\cos2 \phi) < 0 }{N((\cos2 \phi) > 0 ) + N((\cos2 \phi) < 0 )} = \frac{2}{\pi}  \frac{2 \lambda_{\phi}}{3+ \lambda_{\theta}}
		\label{population2}
	\end{equation}

	\begin{eqnarray}
		&& \frac{
			N(\sin(2\theta)\cos\phi > 0) - N(\sin(2\theta)\cos\phi < 0)
		}{
			N(\sin(2\theta)\cos2\phi > 0) + N(\sin(2\theta)\cos2\phi < 0)
		} \nonumber\\  
		&&\hspace{4.5cm} = \frac{2\lambda_{\theta \phi}}{3 + \lambda_{\theta}}
		\label{population3}
	\end{eqnarray}
	
	{The advantage of this method is that it is simple and less dependent on fitting assumptions but limited precision due to discrete bin counting and may not fully account for detector acceptance effects.} \\
	
	\item Average angular distribution method: \\
	
	The average angular distribution method relies on computing mean values of trigonometric functions of the decay angles, such as 
	$\langle \cos^{2}\theta \rangle$, 
	$\langle \sin^{2}\theta \cos 2\phi \rangle$, 
	and $\langle \sin 2\theta \cos \phi \rangle$, 
	which are directly related to the polarization parameters:

	In this method $\lambda$ polarization parameters are estimated following the Eqs. [\ref{av_ang1}, \ref{av_ang2} $\&$ \ref{av_ang3}] given below:
	\begin{eqnarray}
		\langle \cos^2\theta \rangle =  \frac{1 + \frac{3}{5}  \lambda_{\theta}}{3 + \lambda_{\theta}}
		\label{av_ang1}
	\end{eqnarray}
	\begin{eqnarray}
		\langle \cos2\phi \rangle =  \frac{ \lambda_{\phi}}{3 + \lambda_{\theta}}
		\label{av_ang2}
	\end{eqnarray}
	\begin{eqnarray}
		\langle \sin2\theta \cos\phi \rangle =  \frac{4}{5} \frac{ \lambda_{\theta\phi}}{3 + \lambda_{\theta}}
		\label{av_ang3}
	\end{eqnarray}
	
	{This method is straightforward, computationally efficient and gets less affected by binning choice but it requires accurate acceptance corrections and may introduce bias if angular acceptance is non-uniform.} \\
	
	These methods offer a different complementary approach to analyze the decay muon angular distribution as compared to the multi/ single - parameter fitting strategy~\cite{DK2069}. In experimental measurements, different procedures are followed to better understand the systematics associated with the extracted polarization parameters. 
\end{enumerate}

\begin{figure}[ht]
	\centering
	\includegraphics[width=1.00\linewidth]{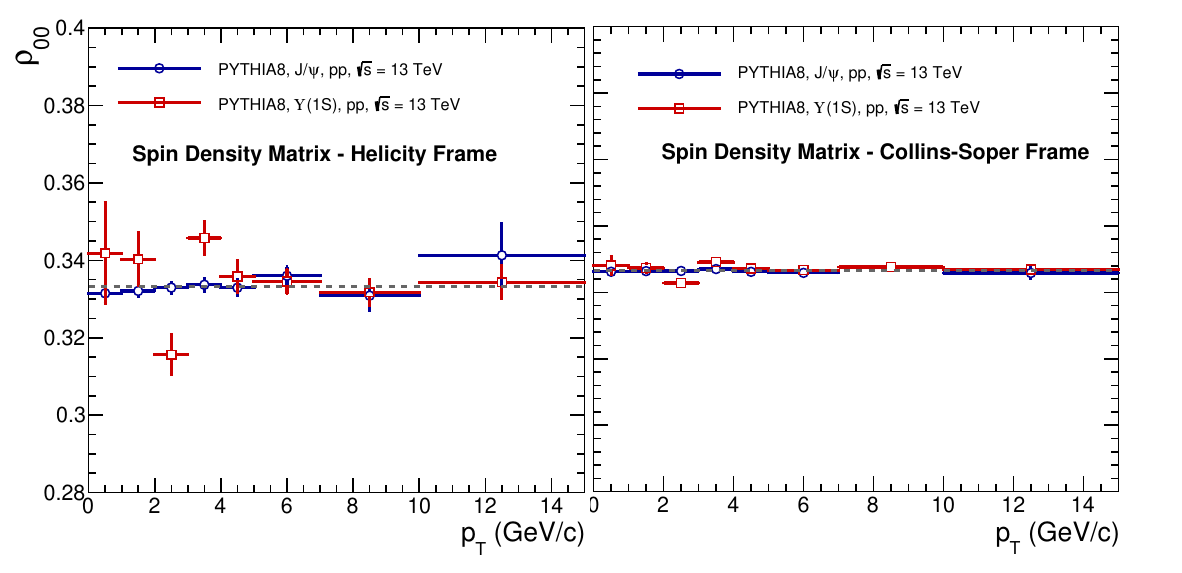}
	\caption{ The spin density matrix element ($\rho_{00}$) as a function of transverse momentum ($p_{T}$) for $J/\psi$ and $\Upsilon(1S)$ in p–p collisions at $\sqrt{s} = 13$ TeV, in the HE  (left) and CS (right) reference frames. }
	\label{rho_pt}
\end{figure}

\section{RESULTS}

\begin{figure}[ht]
	\centering
	\includegraphics[width=1.00\linewidth]{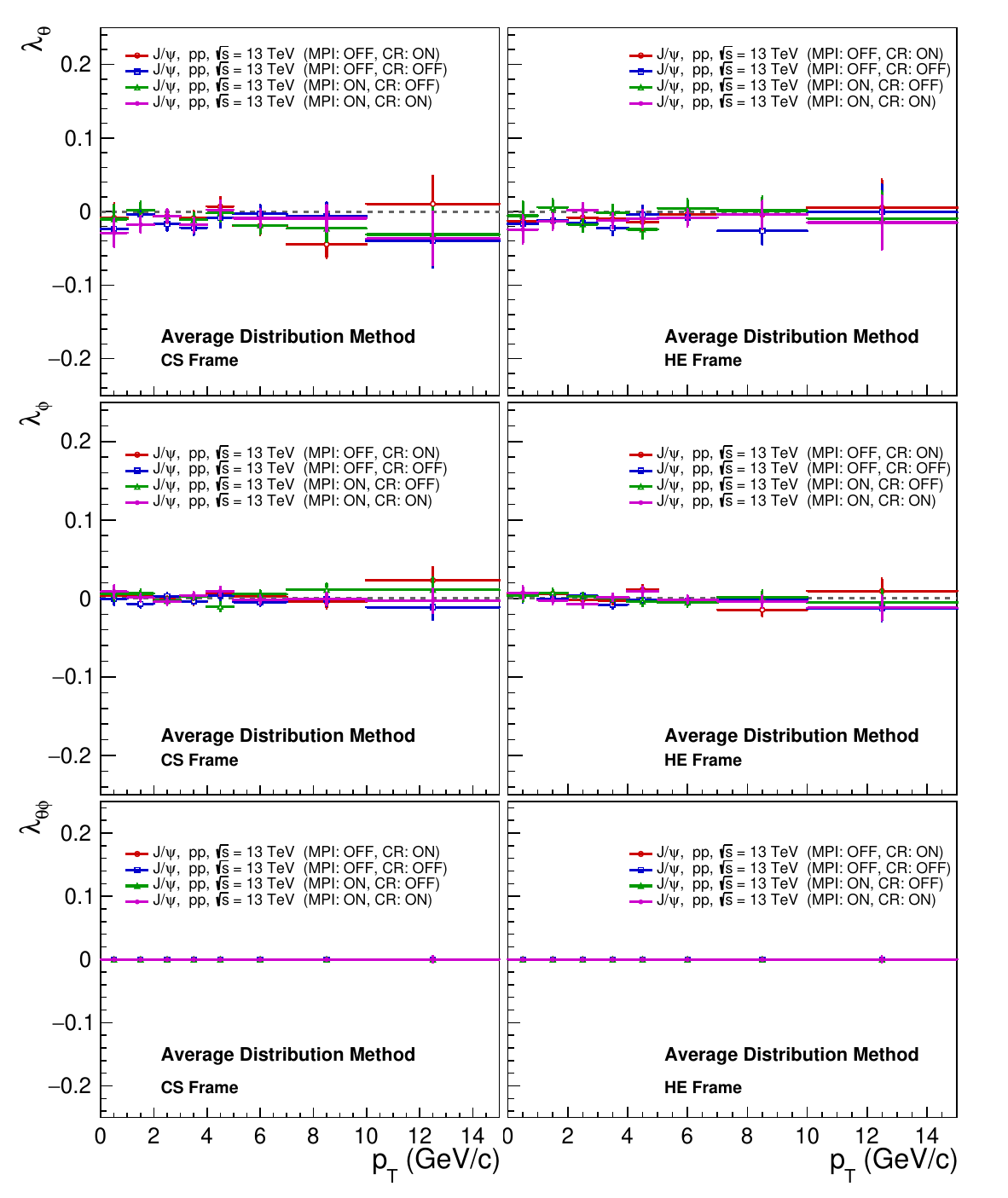}
	\caption{ The polarization parameters ($\lambda_{\theta}$, $\lambda_{\phi}$, and $\lambda_{\theta\phi}$) as functions of transverse momentum ($p_T$) for $J/\psi$  in p–p collisions at $\sqrt{s} = 13$ TeV, as simulated with PYTHIA8 under different parton-level settings. The results are shown in both  HE and CS frames using the average distribution method. Comparisons are made between simulations with MPI and CR both turned off, MPI turned on but CR off, MPI turned off but CR on and both MPI and CR turned on, to investigate the impact of multi-parton interactions and color reconnection on the extracted polarization parameters. }
	\label{Jpsi13_MPI_CR}
\end{figure}

As a first step toward extracting quarkonium polarization parameters; \(\lambda_\theta\), \(\lambda_\varphi\), and \(\lambda_{\theta\varphi}\), we compute the cosine of the polar angle (\(\cos\theta\)) and the azimuthal angle (\(\varphi\)) of the decay muons originating from quarkonium decays, analyzed in both the CS and HE reference frames. To ensure consistency with procedures employed in experimental analyses, we enforce the decay of quarkonium states in di-muon channel (e.g., \(J/\psi \to \mu^+ \mu^-\)) and reconstruct them via the invariant mass of the decayed muons pairs.

Although our simulation yields a delta-function-like invariant mass distribution centered at the quarkonium pole mass---due to the absence of detector smearing effects---we nonetheless follow the similar invariant mass reconstruction procedure to ensure compatibility when comparing our results to those from experimental measurements.
In particular, since our study aims to compare with the ALICE experiment at forward rapidity, we apply a kinematic selection on the reconstructed dimuon rapidity, restricting it to the range \(2.5 < y < 4.0\), in accordance with ALICE Muon Spectrometer (MS) acceptance.

\begin{figure}
	\includegraphics[width=1.00\linewidth]{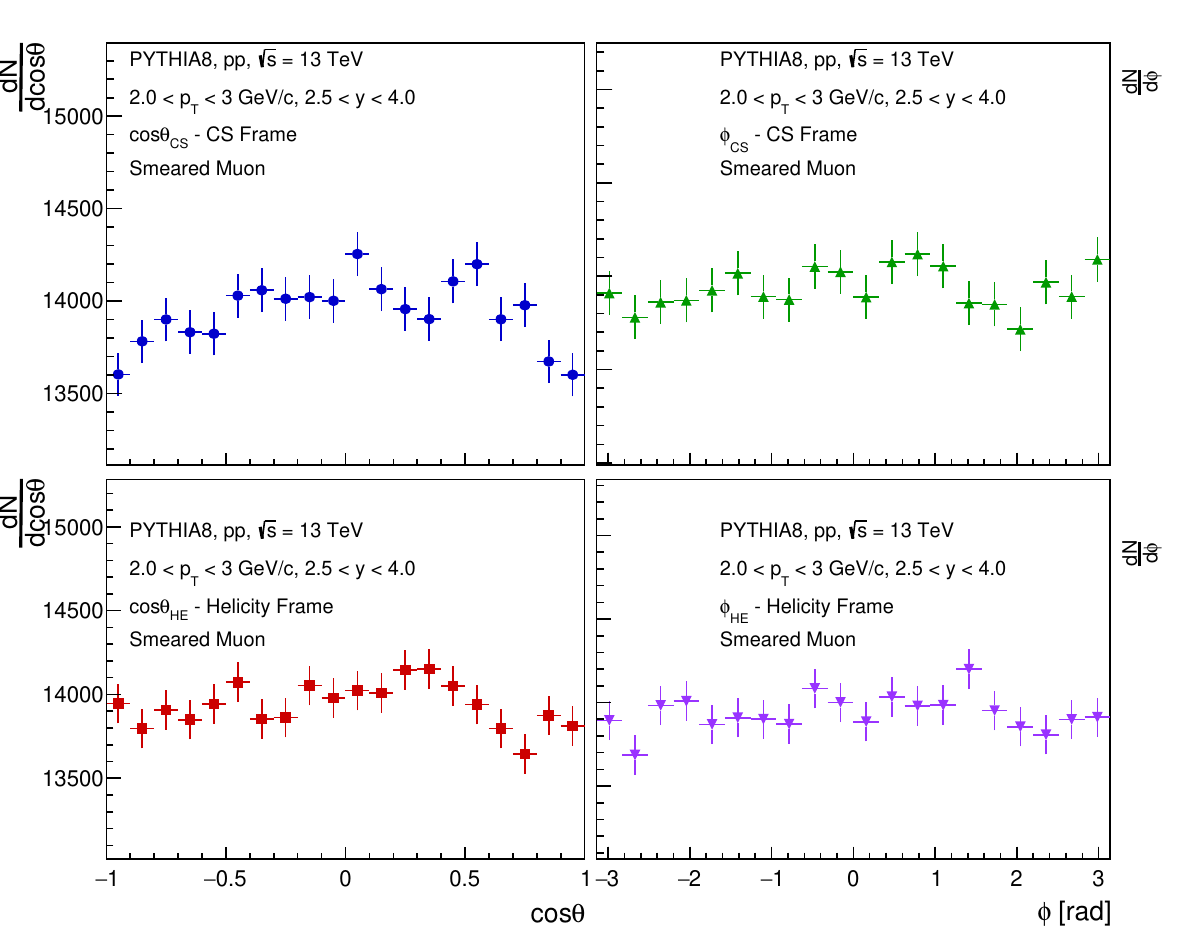}
	\includegraphics[width=1.00\linewidth]{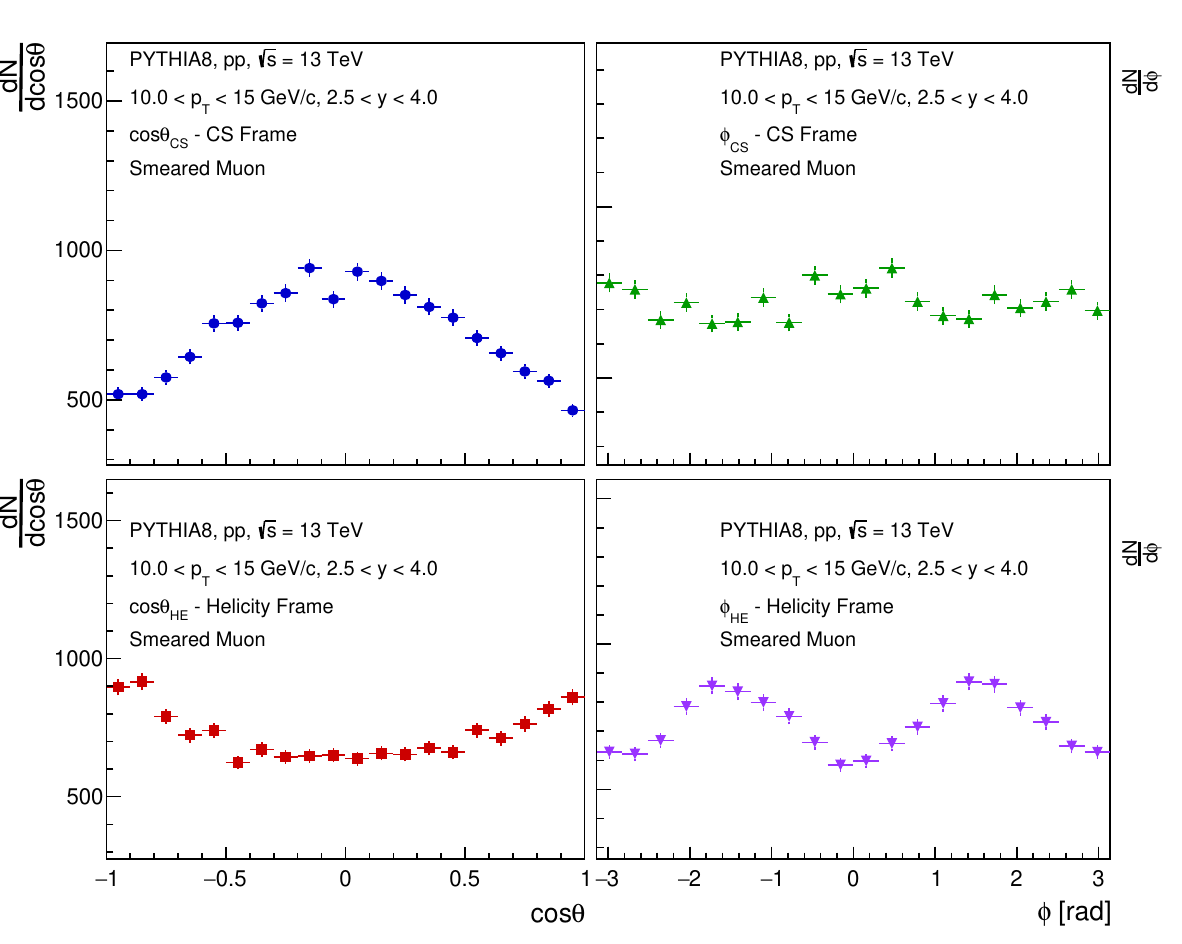}
	\caption{Angular distributions of smeared muons from $J/\psi \to \mu^+\mu^-$ decays in p–p collisions at $\sqrt{s} = 13$ TeV. The distributions are shown for the transverse momentum interval $2 < p_T < 3$ GeV/$c$,  $10 < p_T < 15$ GeV/$c$ and rapidity $2.5 < y < 4.0$. Top panels show the $\cos\theta$ and $\phi$ distributions in the CS frame, while the bottom panels correspond to the HE frame. These distributions are used to extract polarization parameters and account for detector smearing effects.}
	
	\label{Angular_plotvsPt_smeared}		
\end{figure}	

Figure~\ref{Angular_plotvsPt} presents the distributions of the polar angle (\(\cos\theta\)) and the azimuthal angle (\(\varphi\)) for \(J/\psi\) mesons in the transverse momentum range \(2.0 < p_{\mathrm{T}} < 3.0~\mathrm{GeV}/c\), as obtained from \texttt{PYTHIA8} simulations with MPI and CR enabled, for pp collisions at \(\sqrt{s} = 13~\mathrm{TeV}\) in both the CS and HE reference frames. In both frames, the distributions appear flat with no significant modulation, indicating that the decay of the \(J/\psi\) mesons is isotropic, consistent with no observable polarization of the quarkonium states within this kinematic range. Note that this is in contrast with the previous investigation~\cite{DK2099}, where anisotropies have been observed in both polar and azimuth distributions with similar PYTHIA8 settings, as used in the present calculations.

\begin{figure}[ht]
\centering
\includegraphics[width=1.00\linewidth]{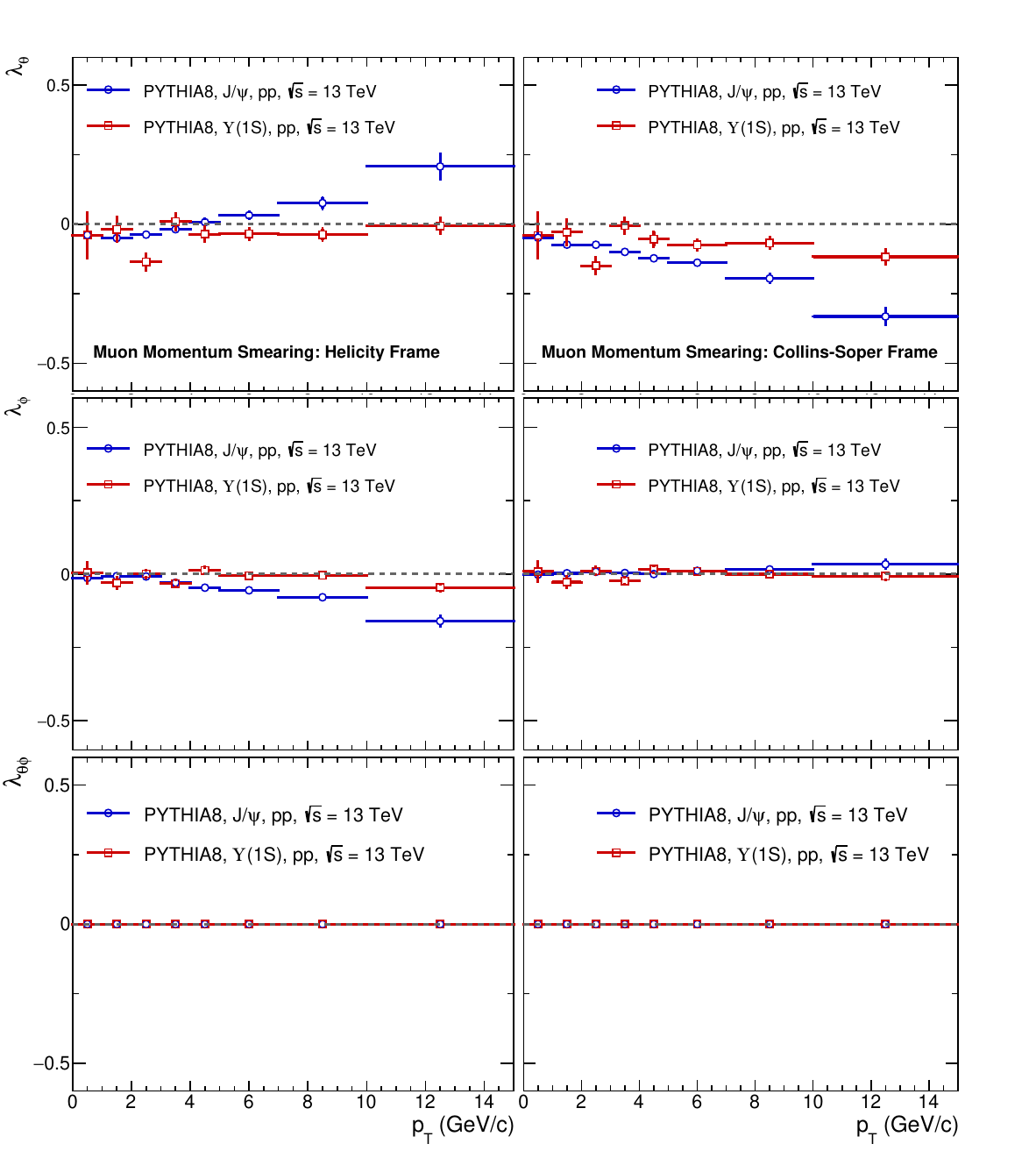}
\caption{ The polarization parameters ($\lambda_{\theta}$, $\lambda_{\phi}$, and $\lambda_{\theta\phi}$) as functions of transverse momentum ($p_T$) for $J/\psi$ and $\Upsilon(1S)$ mesons in p–p collisions at $\sqrt{s} = 13$ TeV. The results are presented in both the HE and CS reference frames. $1 \%$ single muon momentum smearing has been applied to mimic detector resolution effects. The figure illustrates the sensitivity of polarization observables to detector-level distortions for both charmonium and bottomonium states.}
\label{smearing_lambda_vs_pt}
\end{figure}

Having obtained the angular distributions, we now proceed to extract the polarization parameter \(\lambda_\theta\) for both the \(J/\psi\) and \(\Upsilon(1S)\) mesons as a function of transverse momentum (\(p_{\mathrm{T}}\)), using three independent methods: the average method, the one-dimensional (1D) fit method, and the population counting method as shown in Fig.~\ref{lam_pt_jpsi_upsilon}. All three approaches consistently indicate no significant polarization, with \(\lambda_\theta\) values statistically compatible with zero across the studied \(p_{\mathrm{T}}\) range.

Based on the extracted values of \(\lambda_\theta\), we further estimate the corresponding spin density matrix element, \(\rho_{00}\), using the relation :

\begin{center}
\(
\lambda_\theta = \frac{1 - 3\rho_{00}}{1 + \rho_{00}}.
\)
\end{center}

In Figure~\ref{rho_pt}, we find \(\rho_{00} = \frac{1}{3}\) across the entire \(p_{\mathrm{T}}\) range, indicating the absence of a preferred spin alignment. This results further confirms the isotropic nature of the angular distributions of the decayed muon pairs observed in our analysis.
These results are also in stark difference with the previous computations~\cite{DK2099}, where $\lambda$ polarization parameters have been obtained via average distribution method. The behavior of $\lambda_{\theta}$ in both reference frames, has shown indication of $J/\psi$ being longitudinally polarized at low $p_{T}$ and transversely polarized at high $p_{T}$. The $\lambda_{\phi}$ parameter on the other has been found to be negative in HE and positive in CS frame for $J/\psi$ mesons across all energies. The estimated values of $\lambda_{\theta\phi}$ parameter has been found to be close to zero as a function of $p_{T}$. Observation of non-zero polarization has been attributed to the underlying physics processes in PYTHIA8 like feed down and final state gluon radiation. Interestingly even though we have generated inclusive data samples for both quarkonia states, we do not see any hint of non-zero polarization.
\begin{figure}[h]
\centering
\includegraphics[width=0.65\linewidth]{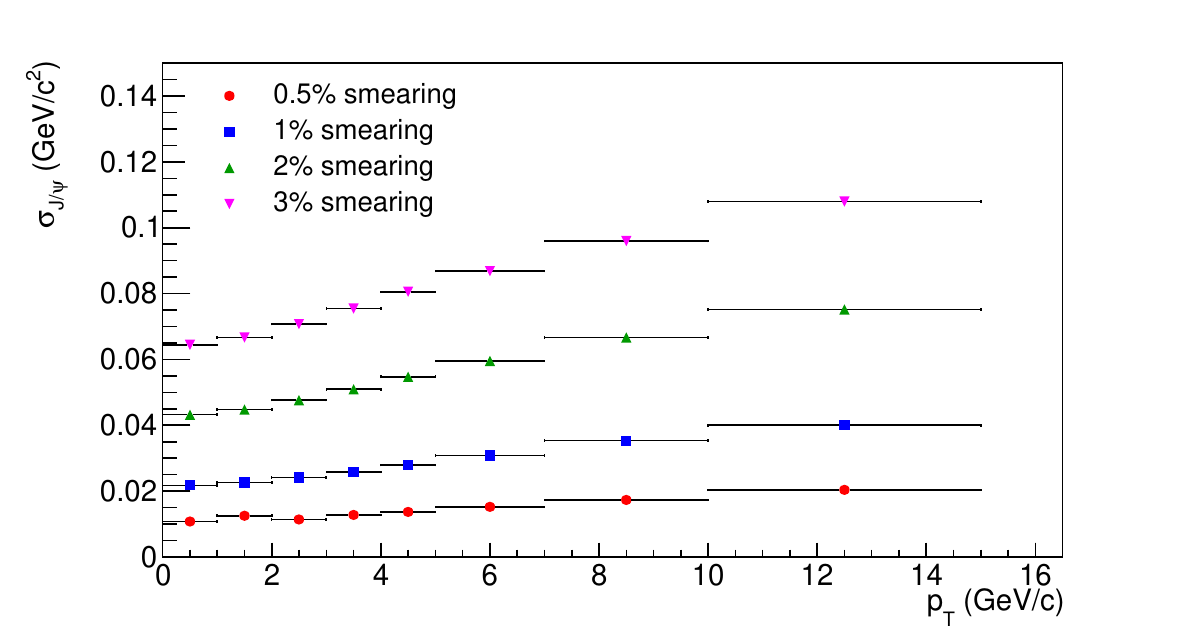}
\caption{ J/$\psi$ mass resolution ($\sigma_{J/\psi}$) as a function of $p_{\rm T}$ for four different values of single momentum smearing conditions. The single muon momentum resolutions are so chosen that the corresponding pair mass resolutions are comparable to values reported by LHCb and ALICE experiments.}
\label{mass_sigma}
\end{figure}

To further investigate whether variations in the underlying physics processes can induce polarization in quarkonium states, we extract the polarization parameters using different configurations of \textsc{Pythia8}, specifically: \texttt{MPI:on \& CR:off}, \texttt{MPI:off \& CR:on} and \texttt{MPI:off \& CR:off}, and compare them with the default setting, i.e., \texttt{MPI:on \& CR:on}. As shown in Fig. ~\ref{Jpsi13_MPI_CR} , all extracted polarization parameters remain consistent with zero across these different settings. This observation suggests that neither MPI nor CR as implemented in \textsc{PYTHIA8}-induces any significant polarization in quarkonium states.

This is rather expected because quarkonia are predominantly produced in the hardest partonic scattering of the event. Subsequent semi-hard or soft parton interactions, such as those modeled by MPI, are unlikely to influence the spin alignment of the quarkonium state directly. As a result, such secondary partonic interactions cannot introduce polarization in quarkonium states. Similarly, color reconnection (CR) is a non-perturbative process that affects hadronization by reconnecting color strings originating from different partonic interactions. Since CR primarily alters the final-state color topology without coupling to the spin degrees of freedom, it is also not expected to introduce any spin alignment in quarkonium production.

\begin{figure}[h]
\centering
\includegraphics[width=1.0\linewidth]{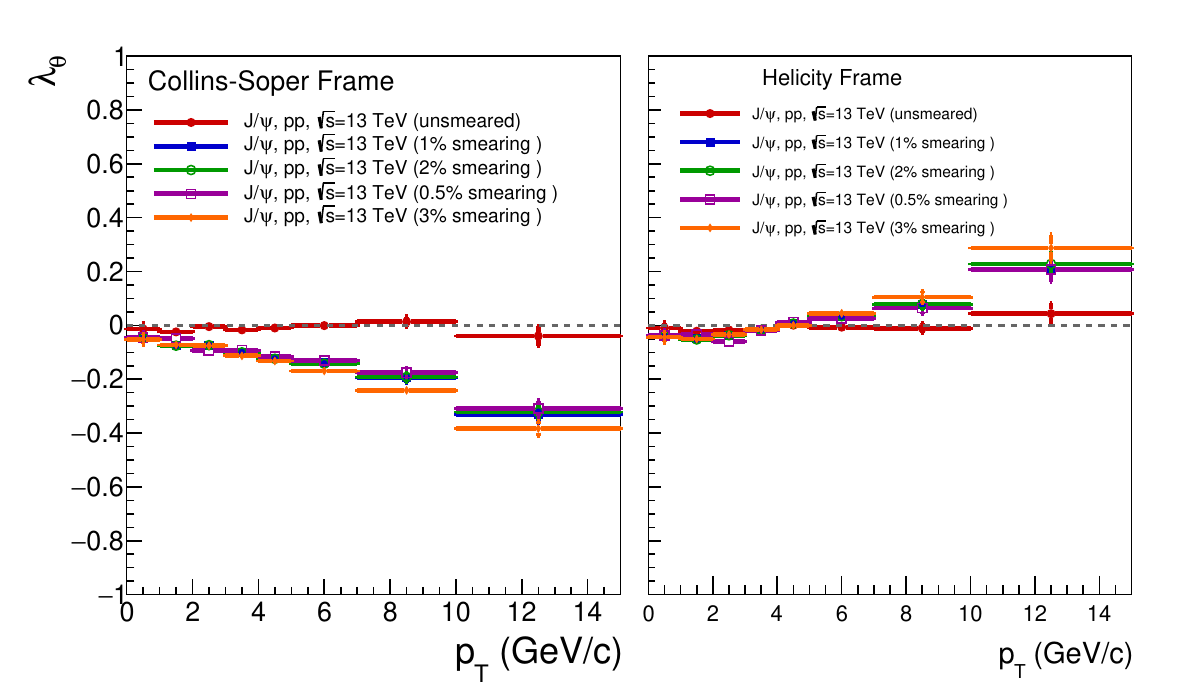}
\caption{The polarization parameter $\lambda^{\rm CS}_{\theta}$ (left) and $\lambda^{\rm HE}_{\theta}$ (right) as a function of $p_{\rm T}$ for different amounts of single muon momentum smearing.}
\label{lambda_CS_HE}
\end{figure}


As highlighted in previous studies, the extraction of polarization parameters is highly sensitive to detector acceptance, efficiency effects, and the associated correction procedures. In particular, the angular distributions of decay products,  can be subjected to significant distortion by the limited geometric acceptance, non-uniform detector response, and kinematic thresholds imposed by the detector system.
If these effects are not properly accounted for, or if residual distortions remain even after correction,
the measured polarization parameters can be biased or misleading. For instance, an isotropic angular distributions of decayed muons at the generator level may appear anisotropic after detector-level smearing and reconstruction, falsely indicating polarization.
Consequently, any meaningful comparison between simulation and experimental data, or between different experiments, requires that detector effects be consistently corrected for.

\begin{figure*}[ht]
\centering
\includegraphics[width=0.75\linewidth]{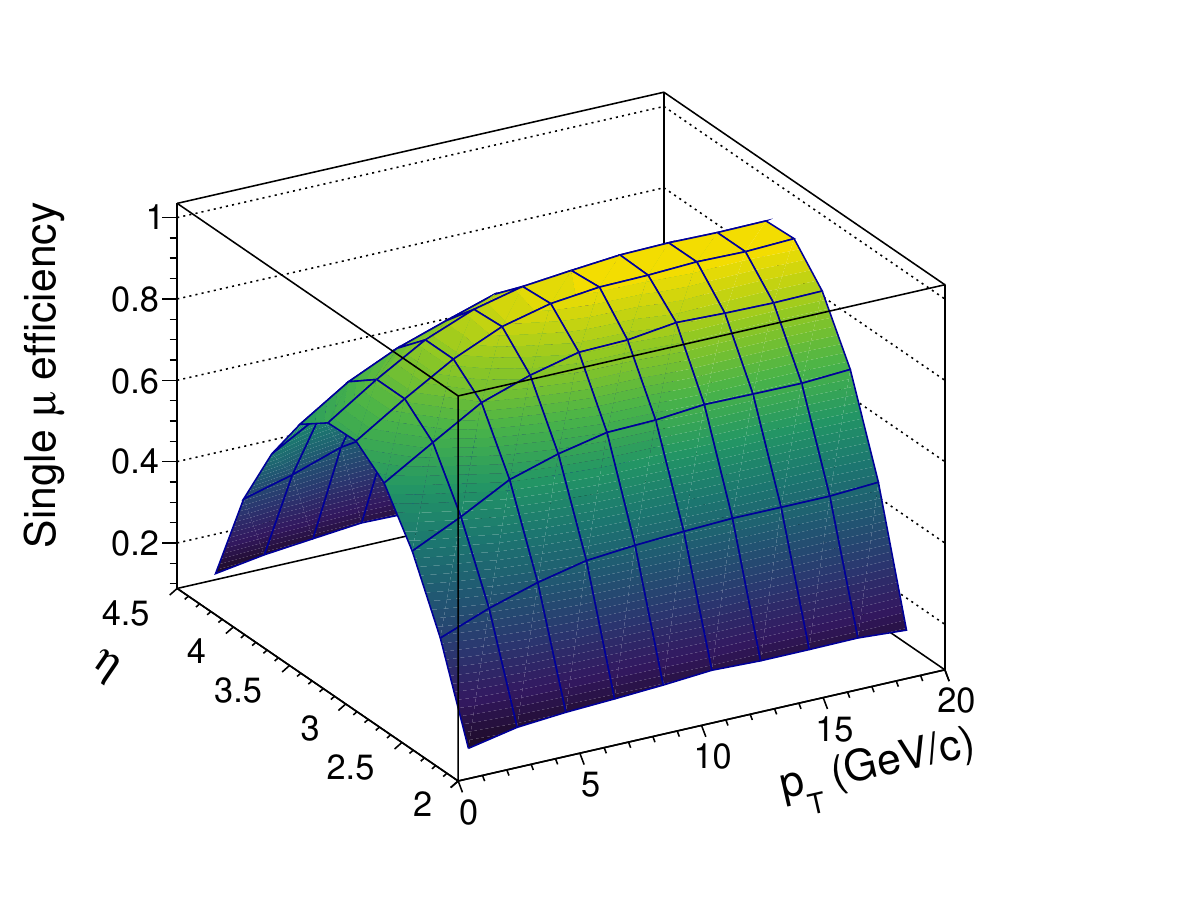}
\caption{Single muon efficiency matrix as a function of p$_{T}$ and $\eta$ used for probabilistically accept or reject muons from quarkonia decay.}
\label{muon_eff_matrix}
\end{figure*}


To address potential biases introduced by detector effects, we simulate the effects of momentum smearing and inefficiencies in muon detection. The former is implemented by applying a Gaussian smearing to the relative momentum resolution, \(\delta p / p\), of the \(p_x\), \(p_y\), and \(p_z\) components of the single muon momentum, with a resolution width of 1\%. For the latter, we construct a single-muon efficiency matrix as a function of  \(p_{\mathrm{T}}\) and \(\eta\). Using this matrix, we apply an acceptance-rejection method in which each generated muon is probabilistically accepted or rejected based on the corresponding efficiency value in its \((p_{\mathrm{T}}, \eta)\)-bin.
Following this, we compute the dimuon efficiency as a function of the decay angular variables, i.e., \(\cos\theta\) and \(\varphi\), in both the CS and HE reference frames.
While the single-muon efficiency is applied over the entire event sample that we analyse but to ensure statistical independence between the efficiency estimation and its application, we divide the full event sample into two disjoint subsets in a 40:60 ratio. The efficiency correction factors are derived from the 40\% subset and subsequently applied to the remaining 60\% of the sample.

\begin{figure}[ht]
\centering
\includegraphics[width=1.00\linewidth]{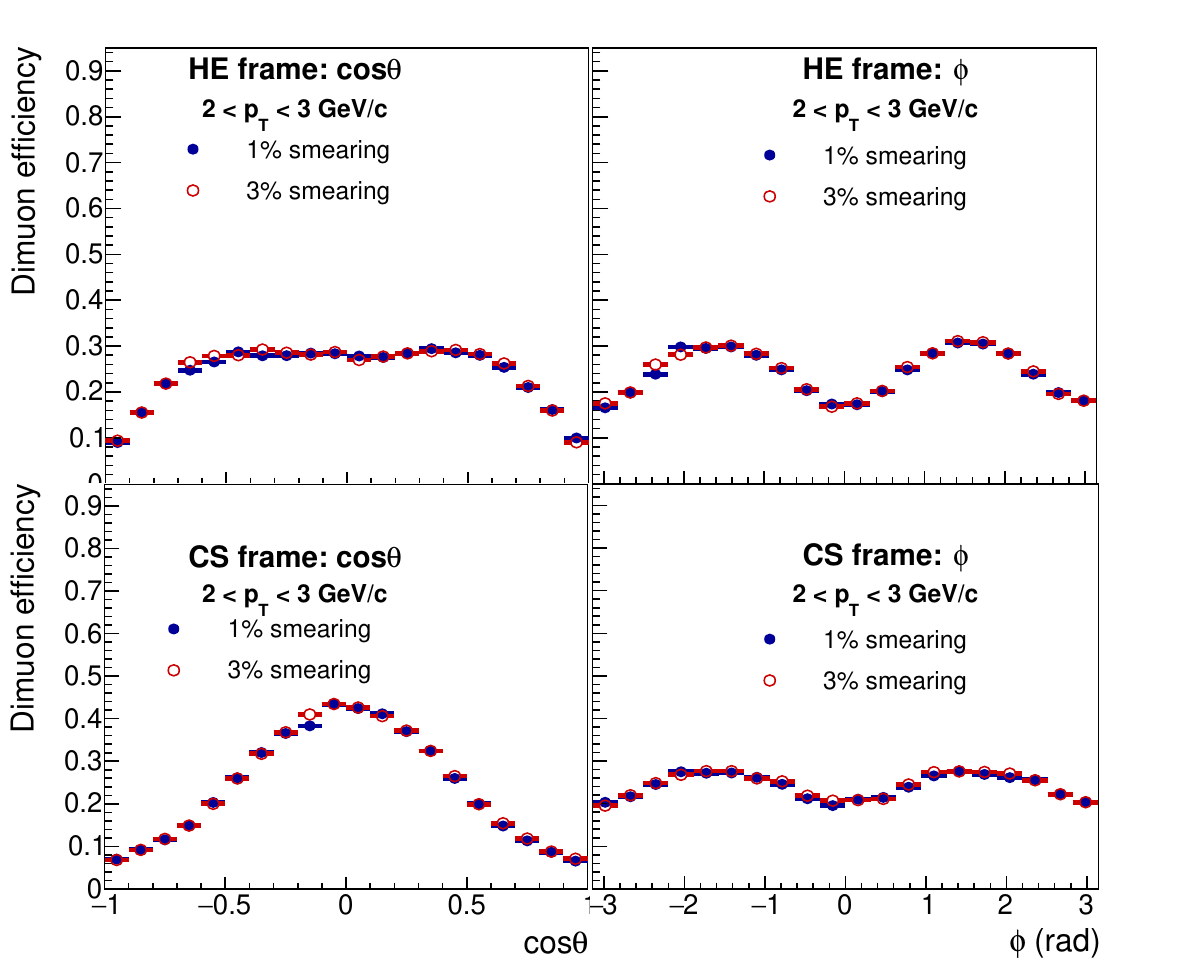}
\caption{Dimuon efficiency as a function of \(\cos\theta\) and \(\varphi\), obtained after including $1\%$ and $3 \%$ single muon momentum smearing and $\pm 3\sigma$ pair mass window cut in 2 $< p_{T} <$ 3 GeV/c.}
\label{dimuon_eff_distributions}
\end{figure}

Upon applying momentum smearing to single muons, the dimuon invariant mass which was originally a delta function centered at the quarkonium pole mass (as given by the PDG), becomes broadened due to the detector resolution effects. The resulting invariant mass distribution is fitted using a double-sided Crystal Ball function to model both the Gaussian core and the non-Gaussian tails. To suppress background and ensure mass consistency with the quarkonium signal, only those dimuon candidates falling within \(\pm 3\sigma\) around the fitted mean mass are retained for further analysis. With this mass window selection imposed, we observe from  Fig.~\ref{Angular_plotvsPt_smeared} that the angular distributions of the decay muons, evaluated in both CS and HE frames, exhibit clear modulations. These modulations mimic the presence of a non-zero polarization, even though the original generator-level sample is unpolarized. This demonstrates that detector resolution effects, coupled with invariant mass selection, can artificially induce angular anisotropies. When extracting the polarization parameters \(\lambda_\theta\), \(\lambda_\varphi\), and \(\lambda_{\theta\varphi}\) from these angular distributions, clear non-zero values for \(\lambda_\theta\) and \(\lambda_\varphi\) are observed as functions of \(p_{\mathrm{T}}\) even in the originally unpolarized samples. This effect is notably more pronounced for the \(J/\psi\) meson compared to the \(\Upsilon(1S)\), reflecting enhanced sensitivity to detector resolution effects at lower quarkonium masses. The underlying reason for this is the relation between muon kinematics and the reconstructed dimuon invariant mass \(M\). Approximating the invariant mass as

\begin{center}
\(
M \approx \sqrt{2\,p_1 p_2 (1 - \cos\alpha)},
\)  
\end{center}

where \(p_i\) are the muon momenta and \(\alpha\) is the opening angle between them, the mass resolution induced by a momentum smearing \(\delta p\) can be approximated by
\begin{center}
\(
\delta M \approx \sqrt{(1 - \cos\alpha)} \cdot \delta p.
\) 
\end{center}

This expression shows that the mass resolution explicitly depends on the decay kinematics, particularly \(\alpha\), which is correlated to the polar angle \(\theta\) of the decay muon ($\mu^{+}$) in the quarkonium rest frame. Applying a fixed mass-window cut, \(\pm 3\sigma\) around the quarkonium mass peak then implicitly imposes a \(\cos\theta\)-dependent selection efficiency that distorts the observed angular distributions. Since the \(J/\psi\) mass (\(\sim 3.1\,\mathrm{GeV}/c^2\)) is significantly smaller than that of the \(\Upsilon(1S)\) (\(\sim 9.46\,\mathrm{GeV}/c^2\)), its decay muons have lower average momentum and consequently exhibit a larger opening angle in the laboratory frame. This leads to a broader reconstructed mass distribution and a stronger angular-dependent acceptance bias from the mass-window selection. Conversely, the heavier \(\Upsilon(1S)\) experiences less mass broadening and reduced distortions in the angular distributions. This interplay between quarkonium mass, momentum resolution, and selection condition explains how detector smearing and mass-window cuts induce more significant artificial polarization effects in the \(J/\psi\) than in the \(\Upsilon(1S)\).
Interestingly, as can be seen from Fig.~\ref{smearing_lambda_vs_pt}, in the HE frame, \(\lambda_\theta\) for \(J/\psi\) exhibits a clear trend: at low transverse momentum (\(p_{\mathrm{T}} < 5~\mathrm{GeV}/c\)), the polarization appears longitudinal, i.e., \(\lambda_\theta < 0\); whereas at higher \(p_{\mathrm{T}}\), the distribution becomes transversely polarized, with \(\lambda_\theta > 0\). This transition is not present at the generator level and arises solely from the combination of momentum smearing and mass-window selection. The corresponding effect of finite single muon momentum resolution on the so called frame invariant quantity ($\lambda_{inv}$) is discussed in Appendix A.
{At this juncture it would be further interesting to investigate the effect of variation in single muon momentum smearing on the corresponding mass resolution of J/$\psi$ and the combined effect of momentum smearing and mass window selection on the degree of induced artificial polarization on the angular distribution of the reconstructed di-muons. 
For this purpose we have varied the single muon momentum smearing between $0.5 \%$ to $3 \%$. The corresponding mass resolutions of $J/\psi$ is shown in Fig.~\ref{mass_sigma} as a function of $p_{T}$. As expected, larger momentum resolution leads to wider di-muon mass distribution with stronger effect with increasing $p_{T}$. It may be noted that J/$\psi$ mass resolution reported by LHCb collaboration is $\sim$ 12 MeV/c$^2$~\cite{LHCb_pp_7TeV} and that measured by  ALICE collaboration is $\sim$ 72 MeV/c$^2$~\cite{ALICE_pp_7TeV,ALICE_pp_5TeV} (for Run2 data). Thus for 0.5\% and 1\% muon momentum smearing conditions the corresponding mass resolution of J/$\psi$ are comparable to the one measured by LHCb while 2\% and 3\% momentum smearing conditions lead to similar mass resolution of J/$\psi$ as measured by ALICE. The resultant effects of momentum smearing and pair mass selection cut on $\lambda^{\rm CS}_{\theta}$ and $\lambda^{\rm HE}_{\theta}$ are illustrated in Fig.~\ref{lambda_CS_HE}. We observe that the effect of momentum smearing on the polarization parameters is large at high $p_T$. The difference between the values of $\lambda^{\rm CS}_{\theta}$ or $\lambda^{\rm HE}_{\theta}$ between the momentum unsmeared and momentum smeared condition increases as we go from low $p_T$ to high $p_T$, maximum difference is $\sim$ 40\% ($\sim$ 30\%) for $\lambda^{\rm CS}_{\theta}$ ($\lambda^{\rm HE}_{\theta}$) in the $p_T$ range 10 $<$ $p_T$ $<$ 15 GeV/$c^2$.} \\

\begin{figure*}[ht]
\includegraphics[width=1.00\linewidth] {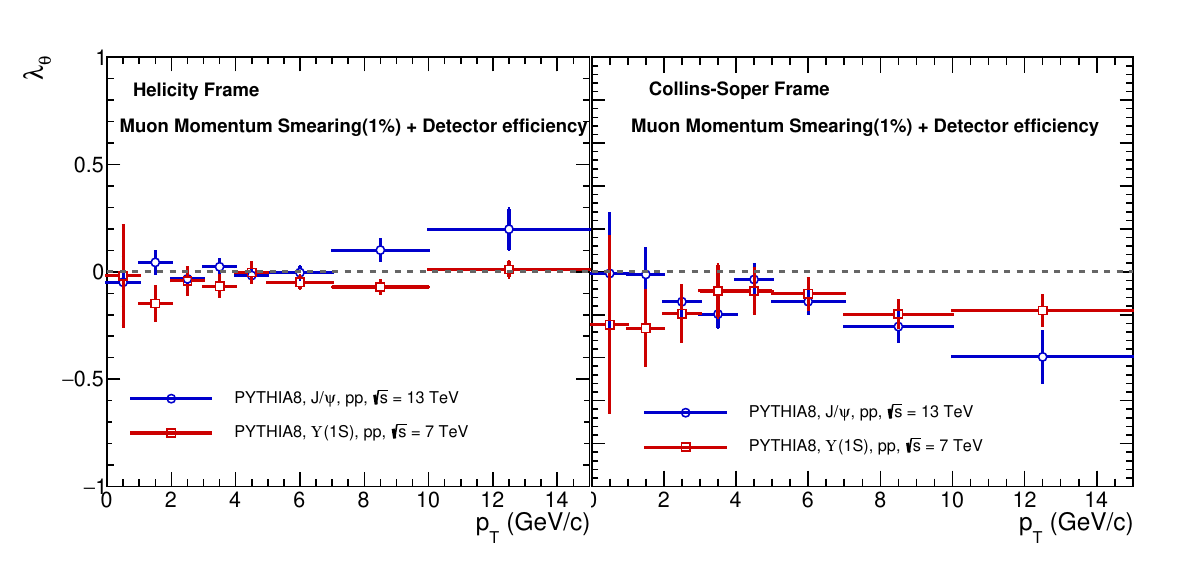}
\includegraphics[width=1.00\linewidth] {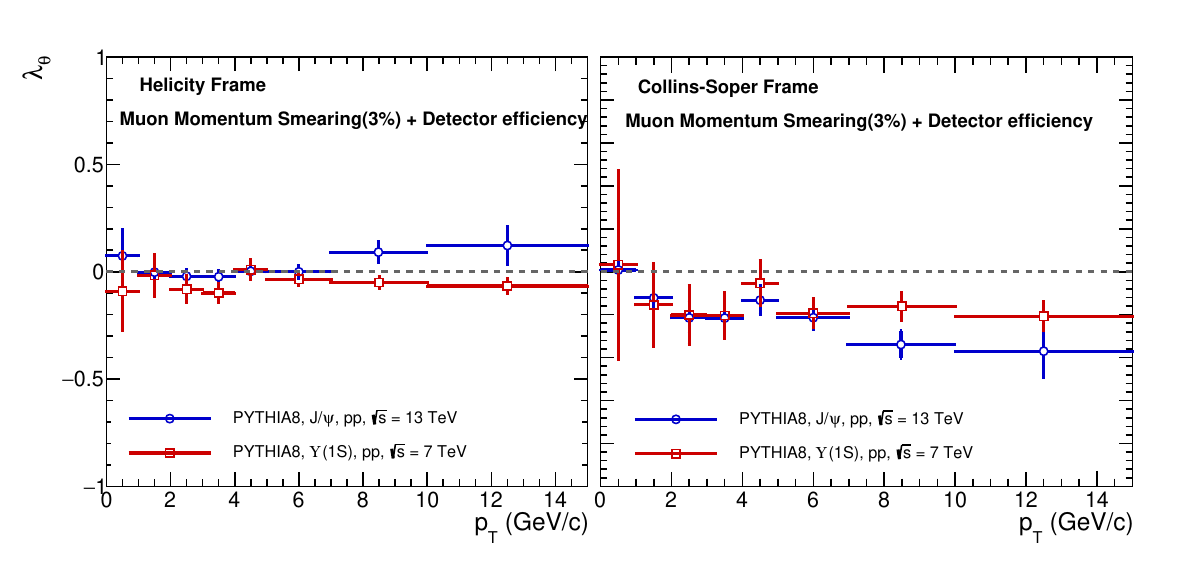}
\caption{The polarization parameter ($\lambda_{\theta}$) as a function of transverse momentum ($p_T$) for $J/\psi$ (red) and $\Upsilon(1S)$ (red) mesons in p–p collisions at $\sqrt{s} = 13$ TeV and 7 TeV, respectively. The results are shown in the HE (left) and CS (right) reference frames. Muon momentum smearing ($1 \%$) and detector efficiency corrections have been applied to mimic realistic experimental effects.}
\label{smearing_efficency_lambda_vs_pt}
\end{figure*}

As previously described, we now introduce detector inefficiencies in addition to momentum smearing and evaluate their combined effect on the dimuon angular efficiency. Specifically, we determine the efficiency as a function of \(\cos\theta\) and \(\varphi\) in both the CS and HE frames, after applying a \(\pm 3\sigma\) mass window cut around the fitted dimuon invariant mass peak, for $1 \%$ and $3 \%$ single muon momentum smearing. To ensure an unbiased correction, the full sample for each case is divided into two statistically independent subsets in a 40:60 ratio. The 40\% subset is used to derive the angular efficiency correction factors, which are then applied to the remaining 60\% of the sample. This allows us to test whether the original unpolarized angular distribution can be recovered after accounting for detector inefficiencies.
\begin{figure}[htbp]
\centering
\includegraphics[width=1.00\linewidth]{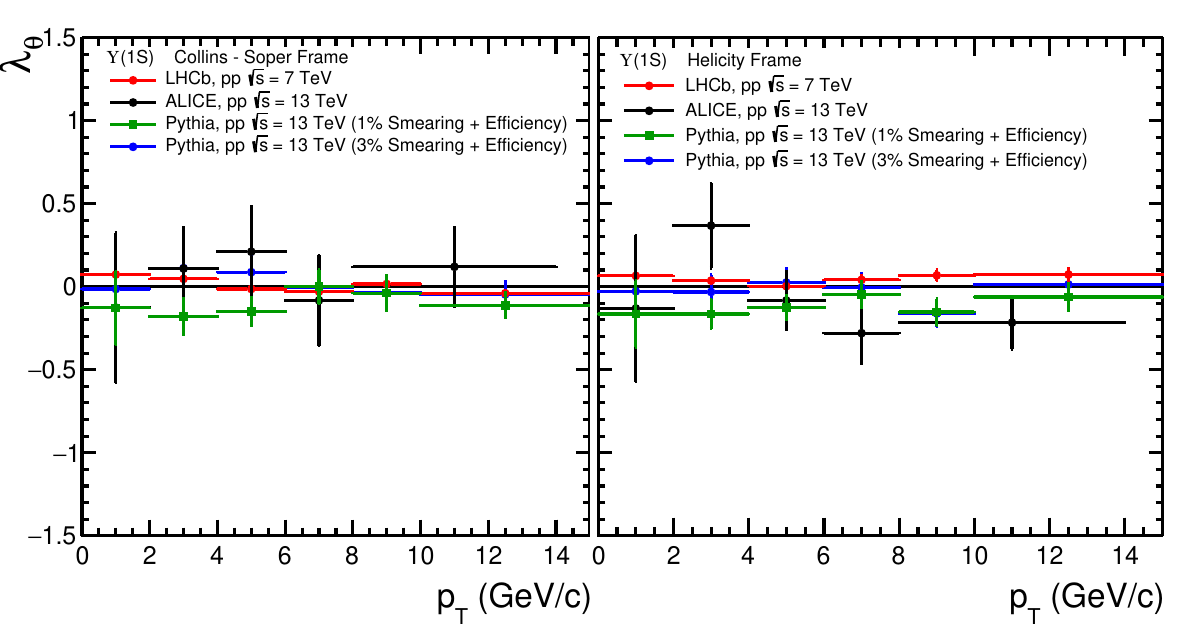}
\caption{Comparison of the polarization parameter $\lambda_\theta$ as a function of transverse momentum $p_T$ for $\Upsilon(1S)$ mesons in proton--proton collisions. Results are shown in the (left) CS, and (right) HE frames. The comparison includes \texttt{PYTHIA8} simulations at $\sqrt{s} = 13$~TeV, corrected for muon momentum smearing ($1 /\%$) and detector effects, experimental data from LHCb at $\sqrt{s} = 7$~TeV, and ALICE data at $\sqrt{s} = 13$~TeV.
}
\label{DrawLamThetaComparison_upsilon_result}
\end{figure}

Figure~\ref{muon_eff_matrix} displays the single-muon efficiency matrix used for the probabilistic acceptance or rejection of generated muons, as a function of \(p_{\mathrm{T}}\) and \(\eta\). The resulting dimuon efficiency distributions, evaluated in terms of \(\cos\theta\) and \(\varphi\) in 2.0 $< p_{T} < ~$3.0 GeV/c, are shown in Figure~\ref{dimuon_eff_distributions}, illustrating the non-uniform acceptance introduced by detector effects across the angular phase space. After correcting for detector inefficiency effects using the derived angular efficiency maps, we find from Fig.~\ref{smearing_efficency_lambda_vs_pt} that the extracted polarization parameters are consistent with zero for both \(J/\psi\) and \(\Upsilon(1S)\) in the CS and HE frames. For $\Upsilon(1S)$ as displayed in Fig.~\ref{DrawLamThetaComparison_upsilon_result}, it agrees well with ALICE preliminary results for $\sqrt{s} = 13$ TeV p-p collisions. No such comparison can be made for $J/\psi$, as the corresponding measurements are still not available.\\
Also to mention, we have repeated the same exercise with a different form of single muon efficiency matrix to understand whether the degree of artificially induced polarization has any explicit dependence on the shape of the single muon efficiency. We find that subsequent to efficiency correction, the qualitative trend of the extracted polarization parameters are almost similar to the default one. Further details can be found in Appendix B.

The above results demonstrate that, within our simulation framework, the applied corrections are effective in recovering the original unpolarized angular distributions. The results are stable with respect to the systematics associated with the momentum smearing and detector inefficiency effects. It should however, be emphasized that our study employs a simplified detector model. In particular, the momentum smearing was assumed to be identical for both the 40\% subset used to derive efficiency maps and the 60\% subset used for applying corrections. In realistic experimental conditions, the detector response may be more complex, introducing residual effects that may not be fully accounted for within this framework. Such residuals can lead to subtle distortions in the angular distributions and potentially generate spurious polarization signals that resemble true physical effects.

\section{Summary and Outlook}

This study investigates the polarization of quarkonium states specifically \(J/\psi\) and \(\Upsilon(1S)\) produced in proton--proton collisions at \(\sqrt{s} = 7\) and \(13~\mathrm{TeV}\), using simulated data generated with \textsc{Pythia8}. The angular distributions of the decay muon pairs are analyzed in both the HE and CS reference frames to extract the polarization parameters \(\lambda_\theta\), \(\lambda_\varphi\), and \(\lambda_{\theta\varphi}\), which characterize the anisotropy of the quarkonium decay angular distributions.

The relevant variables for extraction of polarization parameters, i.e, \(\cos\theta\) and \(\varphi\), are computed following standard experimental procedures. A rapidity selection of \(2.5 < y < 4.0\) is imposed  on dimuon pairs to emulate the acceptance of the ALICE Muon Spectrometer. Polarization parameters are extracted using three independent methods angular averaging, one-dimensional (1D) fitting, and population counting. All three approaches yield results consistent with zero polarization at the generator level, which aligns with the fact that, by construction, quarkonium states produced in \textsc{PYTHIA8} are unpolarized and decay isotropically. Moreover, variations in the underlying event configuration,  enabling or disabling of multiparton interactions (MPI) and color reconnection (CR) in particular, do not significantly affect the extracted polarization parameters. This suggests that MPI and CR, as modeled in \textsc{PYTHIA8}, do not induce spin alignment in quarkonium production.

When detector-level effects are introduced, including muon momentum smearing, acceptance-efficiency corrections, and invariant mass window selections, clear distortions appear in the angular distributions. These lead to artificially non-zero polarization parameters, even for initially unpolarized samples. The effect is particularly significant for the lighter \(J/\psi\) meson, which is more susceptible to distortions due to its lower mass and larger average decay opening angle of muon pairs compared to \(\Upsilon(1S)\). Notably, in the HE frame, the combined influence of smearing and selection biases causes \(\lambda_\theta\) to exhibit a transition from apparent longitudinal polarization at low \(p_{\mathrm{T}}\) to transverse polarization at higher \(p_{\mathrm{T}}\).
After applying acceptance and resolution corrections, the extracted polarization parameters become consistent with zero, thus restoring the expected isotropic decay distributions. For \(\Upsilon(1S)\), the corrected simulation results are in good agreement with preliminary measurements reported by the ALICE experiment at forward rapidity. This study highlights the crucial role of detector effects in quarkonium polarization measurements and the importance of rigorous correction procedures. It demonstrates that even an unpolarized input sample can yield spurious polarization signals if detector resolution and selection biases are not properly accounted for. These findings emphasize the necessity of detailed detector modeling and correction strategies in both experimental and theoretical polarization analyses to avoid potential misinterpretations of the measured polarization parameters. \\

{\bf Data Availability Statement:} Data associated with the manuscript will be made available on reasonable request..



\vspace{2cm}

\appendix*
\section{Frame Invariant Measurement}

It is well established that the polarization parameters \(\lambda_\theta\), \(\lambda_\varphi\), and \(\lambda_{\theta\varphi}\), extracted from the angular distribution of quarkonium decays, are frame-dependent i.e., their values can vary significantly depending on the choice of the quantization axis. This frame dependence complicates direct comparisons across experiments, or between theoretical predictions and experimental measurements, unless the same reference frame is adopted.

To address this ambiguity, it is often recommended to evaluate the frame-invariant polarization parameter \({\lambda}_{inv}\), defined as 

\begin{center}
\(
\lambda_{\mathrm{inv}} = \frac{\lambda_{\theta} + 3\lambda_{\varphi}}{1 - \lambda_{\varphi}}
\)
\end{center}

which provides a more robust characterization of the quarkonium spin alignment, independent of the specific coordinate system.

\begin{figure}[htbp]
\centering
\includegraphics[width=1.00\linewidth]{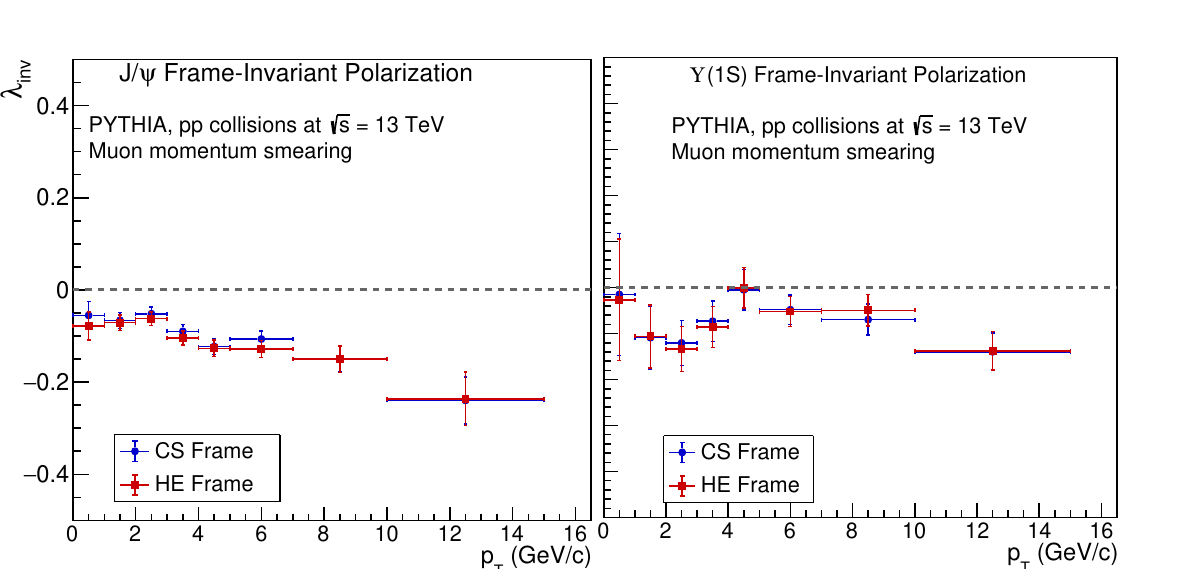}
\caption{Comparison of the frame invariant polarization parameter,\(\tilde{\lambda}_{inv}\),  as a function of transverse momentum ($p_T$) for (left) J/$\psi$  and (right) $\Upsilon(1S)$ mesons, with $1 \%$ single muon momentum smearing, in p–p collisions at $\sqrt{s} = 13$ TeV from \texttt{PYTHIA8}. }
\label{Lambda-invariant}
\end{figure}

As already discussed and shown in Fig.~\ref{smearing_lambda_vs_pt} that the application of muon momentum smearing leads to non-zero values of the polarization parameters, particularly for \(\lambda_\theta\) and \(\lambda_\varphi\), even though the underlying sample remains unpolarized. These non-zero values are therefore spurious and originate from detector resolution effects and kinematic selection biases. 
To evaluate the consistency of the observed polarization across different frames, we compute the frame-invariant parameter \(\lambda_{inv}\), as defined above, with $1 \%$ single muon momentum resolution. We find from Fig.~\ref{Lambda-invariant}, despite the apparent frame-dependent distortions in the individual polarization parameters, the values of \(\lambda_{inv}\) obtained in both the CS and HE frames are consistent with each other within statistical uncertainties. Similar feature is obtained if $3 \%$ single muon momentum resolution is assumed. This confirms the robustness of \(\lambda_{inv}\) as a frame-independent observable and underscores its importance while comparing polarization results across different coordinate systems or experimental setups, particularly when detector effects are present.

\section{Sensitivity of extracted polarization parameters on the efficiency matrix}

\begin{figure}
\centering
\includegraphics[width=0.70\linewidth]{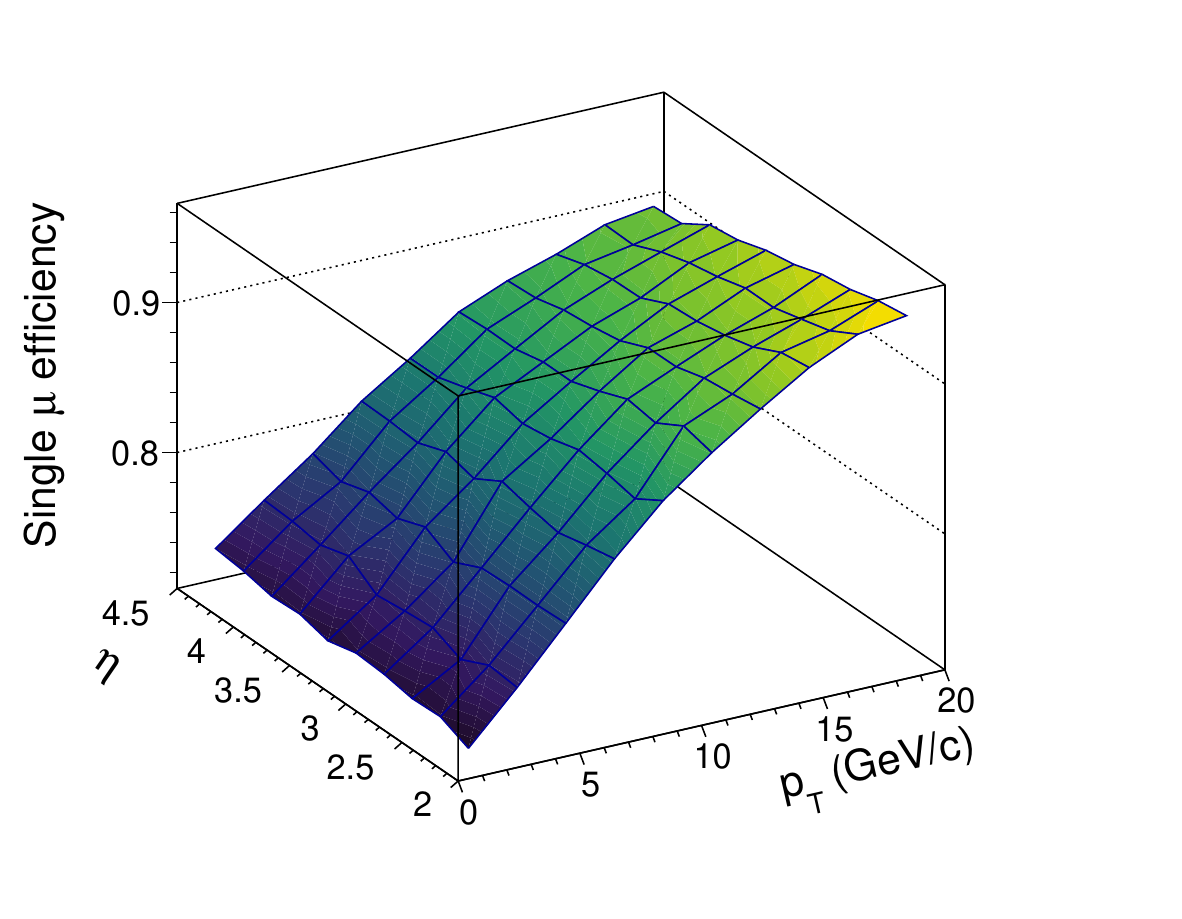}
\caption{A different form of single muon efficiency matrix as a function of p$_{T}$ and $\eta$ used to probabilistically accept or reject muons from quarkonia decay.}
\label{muon_eff_matrix_new}
\end{figure}

\begin{figure*}[hb]
\centering
\includegraphics[width=0.8\linewidth] {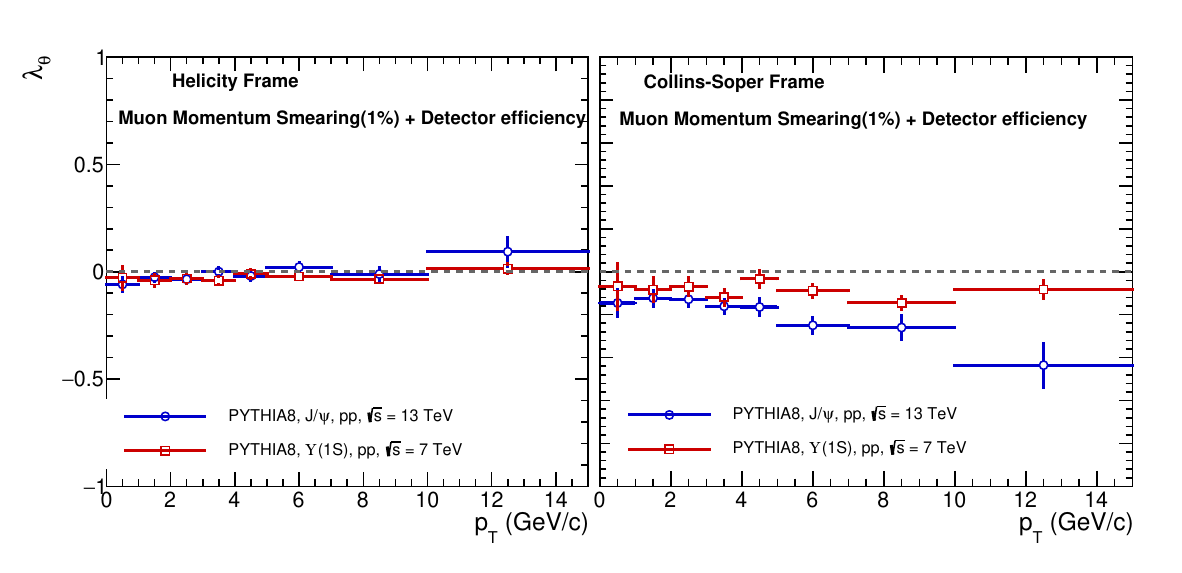}
\includegraphics[width=0.8\linewidth] {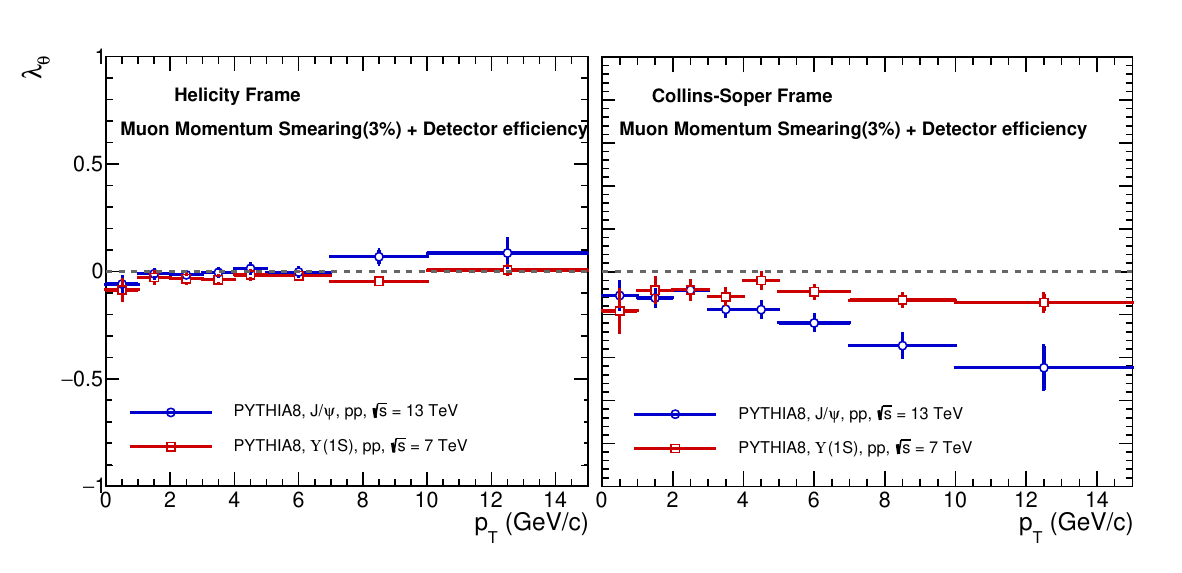}
\caption{The polarization parameter ($\lambda_{\theta}$) as a function of transverse momentum ($p_T$) for $J/\psi$ (red) and $\Upsilon(1S)$ (red) mesons in p–p collisions at $\sqrt{s} = 13$ TeV and 7 TeV, respectively. The results are shown in the HE (left) and CS (right) reference frames. Muon momentum smearing ($1 \%$ and $3\%$) and detector efficiency corrections, following the shape described in Fig.~\ref{muon_eff_matrix_new} have been applied to mimic realistic experimental effects.}
\label{smearing_efficency_lambda_new}
\end{figure*}

We also check the sensitivity of our obtained results on the variation in the shape of the single muon efficiency distribution in the ($\eta$, $p_{T}$) space. For this purpose we choose a different form of the single muon efficiency matrix, that falls steeply to zero at the edges as shown in Fig.~\ref{muon_eff_matrix_new}. With this new efficiency map and following the same strategy as discussed above, we extract the $\lambda_{\theta}$ parameters in CS and HE frames for two characteristic values ($1\%$ and $3\%$) of momentum resolution, as shown in Fig.~\ref{smearing_efficency_lambda_new}. In both the frames results are in line with the default choice.

\end{document}